\documentclass[twocolumn,showpacs,aps,prl,preprintnumbers,amsmath,amssymb,nofootinbib]{revtex4}

\usepackage{latexsym}
\usepackage{graphicx}
\usepackage{dcolumn}
\usepackage{bm}

\newcommand{\be}{\begin{equation}}   
\newcommand{\ee}{\end{equation}}   
\newcommand{\ba}{\begin{eqnarray}}   
\newcommand{\ea}{\end{eqnarray}}  

\begin{document}

\preprint{BI-TP 2006/19, CERN-PH-TH/2006-092, DESY-06-070, MKPH-T-06-10, IFIC/06-17}

\title {\boldmath $K\rightarrow\pi\pi$ amplitudes from lattice QCD with a light charm quark}

\author{L.~Giusti$^a$, P.~Hern\'andez$^b$, M. Laine$^c$, C.~Pena$^a$, 
        J.~Wennekers$^d$, H.~Wittig$^e$}

\affiliation{\vspace{0.3cm}
\hspace{-0.25cm}$^a$ CERN, Department of Physics, TH Division, CH-1211 Geneva 23, Switzerland\\
$^b$ Departamento de Fisica Te\'orica and IFIC, Universidad de Valencia, E-46071 Valencia, Spain\\
\hspace{-0.25cm}$^c$ Faculty of Physics, University of Bielefeld, D-33501 Bielefeld, Germany\\
$^d$ DESY, Notkestra{\ss}e 85, D-22603 Hamburg, Germany\\
$^e$ Institut f\"ur Kernphysik, Universit\"at Mainz, D-55099 Mainz, Germany
}
\date{\vspace{0.2cm} \today}

\begin{abstract}
We compute the leading-order low-energy constants of the $\Delta S=1$ effective weak 
Hamiltonian in the quenched approximation of QCD with up, down, strange, and charm 
quarks degenerate and light. 
They are extracted by comparing the predictions of finite volume chiral 
perturbation theory with lattice QCD computations of suitable correlation functions 
carried out with quark masses ranging from a few MeV up to half of the physical 
strange mass. We observe a $\Delta I=1/2$ enhancement in this corner of the 
parameter space of the theory. Although matching with the experimental result is not 
observed for the $\Delta I=1/2$ amplitude, our computation 
suggests large QCD contributions to the physical $\Delta I=1/2$ rule in the GIM limit, 
and represents the first step to quantify the r\^ole of the charm quark-mass in 
$K \rightarrow\pi\pi$ amplitudes. The use of fermions with an exact chiral symmetry is 
an essential ingredient in our computation.
\end{abstract}


\maketitle

\section{Introduction\label{sec:intro}}
\vspace{-0.25cm}

The decay of a neutral kaon into a pair of pions in a state 
with isospin $I$ is described by the transition amplitudes
\be
i A_I e^{i\delta_I} = 
\langle (\pi\pi)_I| H_\mathrm{w}|K^0\rangle, \qquad I=0,\,2,
\ee
where $H_\mathrm{w}$ is the $\Delta S=1$ effective weak 
Hamiltonian and $\delta_I$ is the $\pi\pi$-scattering 
phase shift. The well-known experimental fact 
\be
   |A_0/A_2| \sim 22\; 
\ee
is often called the $\Delta I = 1/2$ rule. Many decades after  
its experimental discovery, it is embarrassing that 
the origin of this enhancement is still not known. 
In the Standard Model 
(SM) a reliable perturbative computation of short-distance Quantum 
Chromodynamics (QCD) 
corrections \cite{Gaillard:1974nj,Altarelli:1974ex,Altarelli:1980fi,Buras:1989xd}, 
together with a naive order-of-magnitude estimate of long-distance contributions, 
would suggest comparable values for $|A_0|$ and $|A_2|$~\cite{Gaillard:1974nj,Altarelli:1974ex}. 
The bulk of the enhancement is thus expected to come from non-perturbative 
QCD contributions, which makes the $\Delta I = 1/2$ rule one of the rare cases 
where an interplay between strong and electro-weak interactions gives an 
opportunity for a refined test of non-perturbative strong dynamics. 

Lattice QCD is the only known technique that allows us to attack the problem from first 
principles and possibly to reveal the origin of the enhancement~\cite{Cabibbo:1983xa,Brower:1984ta}. 
It would be 
interesting to understand whether it is the result of an accumulation 
of several effects, each giving a moderate contribution, or if it is driven by a 
dominant mechanism. Recently we proposed a theoretically well defined 
strategy to disentangle non-perturbative QCD contributions from the various 
sources~\cite{Giusti:2004an}, and in particular to reveal the r\^ole of the charm quark 
and its associated mass scale (whose relevance in this problem was pointed
out in Refs.~\cite{Shifman:1975tn,Georgi:1985kw}). The main idea is to compute 
the leading-order low-energy constants (LECs)
of the CP-conserving $\Delta S=1$ weak Hamiltonian of the chiral 
low-energy effective theory as a function of the charm quark mass. 
They can be extracted by comparing finite-volume chiral perturbation theory (ChPT) 
predictions for suitable two- and three-point correlation functions with the analogous 
ones computed in lattice QCD at small light-quark masses and momenta. 
The suggestion of using ChPT in connection with kaon amplitudes was pointed out long 
ago \cite{Donoghue:1982cq,Bernard:1985wf}. It is only now that these 
ideas can be formulated and integrated in a well defined strategy~\cite{Giusti:2004an}, 
following significant conceptual advances in the discretization of fermions on the 
lattice as well as enormous gains in computer power. The main theoretical advance is the 
discovery of Ginsparg--Wilson (GW) regularizations 
\cite{Ginsparg:1981bj,Neuberger:1997fp,Hasenfratz:1997ft}, which preserve 
an exact chiral symmetry on the lattice at finite lattice 
spacings \cite{Luscher:1998pq}. By using these fermions the problem 
of ultraviolet power divergences in the effective Hamiltonian $H_\mathrm{w}$ \cite{Maiani:1986db} 
is avoided in the case of an active charm~\cite{Capitani:2000bm}, 
and quark masses as low as a few MeV can be simulated. 
Eventually the full $K\rightarrow\pi\pi$ amplitudes can be computed 
using finite-volume techniques~\cite{Lellouch:2000pv,Lin:2001ek}. 

The aim of this letter is to report on a computation of the LECs of the 
CP-conserving $\Delta S=1$ weak Hamiltonian with up, down, strange, and 
charm quarks degenerate and chiral (GIM limit), i.e. the implementation of the first step 
of the strategy proposed in Ref.~\cite{Giusti:2004an}. 
We perform the first quenched lattice QCD computation of the relevant 
three-point functions with quark masses as light as a few MeV, 
which turns out to be essential for a robust extrapolation 
to the chiral limit. Our results reveal
a clear hierarchy between the low-energy constants, which in 
turn implies the presence of a $\Delta I=1/2$ enhancement in 
this corner of the parameter space of (quenched) QCD. 

Since we are looking for an order-of-magnitude effect, 
and since simulations with dynamical fermions are 
very expensive, it is appropriate for us to first perform 
the computation in quenched QCD. The latter is not a systematic 
approximation of the full theory\footnote{On the other hand 
the ambiguity in the definition of the LECs pointed out by the 
Golterman and Pallante \cite{Golterman:2006ed} is not present in the GIM limit.}. 
However, when quenched results can be compared with 
experimental measurements, discrepancies of ${\cal O}(10\,\%)$ are found
in most cases~\cite{Aoki:2002fd}. 
In the past there were several attempts to attack the problem by using  
quenched lattice 
QCD~\cite{Bernard:1985tm,Gavela:1988ws,Pekurovsky:1998jd,Blum:2001xb,Noaki:2001un,Boucaud:2004aa}. 
In particular, in Refs.~\cite{Blum:2001xb,Noaki:2001un}, a fermion 
action with an approximate chiral symmetry was used and, despite  
the fact that the charm was integrated out and therefore 
an ultraviolet power-divergent subtraction was needed, 
the authors observed a good statistical signal for the subtracted 
matrix elements in a range of quark masses of about half 
the physical strange quark-mass. Several computations of $A_I$ which 
use models to quantify 
QCD non-perturbative contributions in these amplitudes 
can also be found in the literature (see Refs.~\cite{Bijnens:1998ee,Hambye:2003cy}  and references therein).
\vspace{-0.25cm}

\section{\boldmath The $\Delta S =1$ effective Hamiltonian \label{sec:Hds1}}
\vspace{-0.25cm}

In the SU(4) degenerate case and with GW fermions, 
the CP-even $\Delta S =1$ effective Hamiltonian 
is \cite{Gaillard:1974nj,Altarelli:1974ex,Giusti:2004an}
\be\label{eq:HwQCD}
H_{\mathrm{w}} = \frac{g^2_w}{4 M_W^2} V^*_{us} V_{ud}
\Big\{k_1^+ {\cal Q}^+_1 + k_1^- {\cal Q}^-_1\Big\}\, ,
\ee
where 
\ba
{\cal Q}_1^{\pm} & = &   {\cal Z}^{\pm}_{11} \Big\{  
(\bar s\gamma_{\mu}P_{-}{\tilde u})(\bar u\gamma_{\mu}P_{-}{\tilde d})\nonumber\\
& & \;\;\;\ \pm \; (\bar s\gamma_{\mu}P_{-}{\tilde d})(\bar u\gamma_{\mu}P_{-}{\tilde u})
- [u\,\rightarrow\,c]   \Big\} \; ,
\ea
and any further unexplained notation in the paper 
can be found in Ref.~\cite{Giusti:2004an}. We are interested in 
the ratios of correlation functions 
\be\label{eq:Rpm}
R^{\pm} (x_0,y_0) = \frac{C_1^\pm(x_0,y_0)}{C(x_0)C(y_0)}\; ,
\ee
where
\ba
   C(x_0) &=& \sum_{\vec x} \left\langle 
        [J_0(x)]_{\alpha\beta}[J_0(0)]_{\beta\alpha}
                           \right\rangle,   \\
   C_1^\pm(x_0,y_0) &=& \sum_{\vec x,\vec y}
   \left\langle [J_0(x)]_{du}\,[{\cal{Q}}_1^\pm(0)]\,[J_0(y)]_{us}
   \right\rangle\; ,
\ea
$[J_\mu]_{\alpha\beta}={\cal Z}_J 
({\bar \psi}_\alpha \gamma_\mu P_- \tilde\psi_\beta)$,
and ${\cal Z}_J$ is the renormalization constant of the local 
left-handed current.

In the chiral effective theory the corresponding 
effective Hamiltonian reads
\be
{\cal H}_{\mathrm{w}} = \frac{g^2_w}{4 M_W^2} V^*_{us} V_{ud}
\Big\{g_1^+ {Q}^+_1 + g_1^- {Q}^-_1\Big\}\, , 
\ee
where, at leading order in momentum expansion,
\ba
{Q}_1^{\pm} & = & {\displaystyle \frac{F^4}{4}} 
\Big\{
(U\partial_\mu U^{\dagger})_{us}(U\partial_\mu U^{\dagger})_{du} \nonumber\\
& \pm & (U\partial_\mu U^{\dagger})_{ds}(U\partial_\mu U^{\dagger})_{uu}
\;\;\;  - [u \rightarrow c]\Big\} \; .
\ea
The complete expressions at the next-to-leading order (NLO) 
can be found in \cite{Kambor:1989tz,Hernandez:preg}. 
In the quenched approximation of QCD, an effective low-energy
chiral theory is formally obtained if an additional expansion 
in $1/N_c$, where $N_c$ is the number of colours, is carried out 
together with the usual one in quark 
masses and momenta~\cite{Bernard:1992mk,Sharpe:1992ft}. 
Here we adopt the pragmatic assumption that quenched ChPT describes 
the low-energy regime of quenched QCD in certain ranges of 
kinematical scales at fixed $N_c$. Correlation functions can
be parametrized in terms of effective coupling constants,            
the latter being defined as the couplings that appear in the         
Lagrangian of the effective theory. For quark masses light enough to be in
the $\epsilon$-regime of quenched 
QCD~\cite{Gasser:1987ah,Gasser:1987zq,Neuberger:1987fd}, 
the ratios corresponding to Eq.~(\ref{eq:Rpm}) in NLO ChPT, in a 
volume $V=T \times L^3$ and at fixed topological charge $\nu$, 
are~\cite{Giusti:2004an,Hernandez:2002ds}
\be\label{eq:Reps}
K^{\pm}_{\nu}(x_0,y_0) = 1 \pm   
\frac{2\, T}{F^2 L^3} \left\{\beta_1 \Big(\frac{L}{T}\Big)^{3/2} - k_{00}\right\},
\ee
where $F$ is the pseudoscalar decay constant in the chiral limit, and 
the shape coefficients $\beta_1$ and $k_{00}$ can be found 
in Ref.~\cite{Giusti:2004an}. Remarkably, the r.h.s. of Eq.~(\ref{eq:Reps})
is determined once $F$ is known,
and it turns out to be independent of $\nu$ and the quark mass. 
When the quark masses are heavier and reach the so-called $p$-regime of QCD,
the corresponding ratios are given by \cite{Hernandez:preg}

\be\label{eq:Rp}\displaystyle
K^{\pm}(x_0,y_0) = 1 \mp 3 \frac{M^2}{(4\pi F)^2}
\log{\Big(\frac{M^2}{\Lambda_\pm^2}\Big)} 
\pm {\cal K}(x_0,y_0) \, ,
\ee
where $M$ is the pseudoscalar meson mass at LO in ChPT, and 
${\cal K}(x_0,y_0)$ accounts for leading-order 
finite-volume effects and can be found in 
Ref.~\cite{Hernandez:preg}. The LECs $g_1^{\pm}$ can be 
extracted by requiring that
\be
g_1^{\pm} K^{\pm}(x_0,y_0) = k_1^{\pm} R^{\pm} (x_0,y_0)
\ee
for values of quark masses, volumes, $x_0$ and $y_0$, where quenched 
ChPT is expected to parametrize well the correlation functions. 
\vspace{-0.5cm}

\section{Lattice computation \label{sec:Lat}}
\vspace{-0.25cm}

The numerical computation is performed by generating gauge 
configurations with the Wilson action and periodic boundary 
conditions by standard Monte Carlo techniques. The topological charge 
and the quark propagators are computed following 
Ref.~\cite{Giusti:2002sm}. The statistical variance 
of the estimates of correlation functions has been reduced 
by implementing a generalization of the low-mode averaging technique 
proposed in \cite{Giusti:2004yp}, which turns out to be 
essential to get a signal for the lighter quark masses.
\begin{table}[t!]
\begin{center}
\setlength{\tabcolsep}{.25pc}
\begin{tabular}{cc@{\hspace{15pt}}cl@{}l@{\hspace{15pt}}l@{}l}
\hline
$am$ & $aM_P$ & $R^{+,\; \mathrm{bare}}$ & \multicolumn{2}{c}{$R^{-,\; \mathrm{bare}}$} & 
\multicolumn{2}{c}{$(R^{+}\cdot R^{-})^\mathrm{bare}$}\\
\hline
\multicolumn{7}{c}{$\epsilon$-regime}\\
\hline
0.002 &     -      & 0.600(43) & 2&.42(13) & 1&.45(15)\\
0.003 &     -      & 0.603(41) & 2&.40(12) & 1&.44(14)\\
\hline
\multicolumn{7}{c}{$p$-regime} \\
\hline
0.020 & 0.1960(28) & 0.654(40) & 2&.20(12) & 1&.44(12) \\
0.030 & 0.2302(25) & 0.691(33) & 1&.93(9)  & 1&.33(9) \\
0.040 & 0.2598(24) & 0.723(31) & 1&.75(8)  & 1&.26(8) \\
0.060 & 0.3110(24) & 0.772(30) & 1&.51(7)  & 1&.17(8) \\
\hline
\end{tabular}
\caption{Results for $aM_P$ and $R^{\pm,\mathrm{bare}}$\label{tab:lattices}
as obtained from 746 and 197 gauge configurations in the $\epsilon$ and
$p$ regimes, respectively.}
\end{center}
\end{table}
The lattice has a bare coupling constant $\beta\equiv6/g_0^2=5.8485$,  
which corresponds to a lattice spacing $a\sim 0.12$~fm, and a volume of 
$V a^{-4} = 16^3 \times 32$. The list of simulated bare quark masses, 
together with the corresponding results for pion masses and 
unrenormalized ratios 
$R^{\pm,\; \mathrm{bare}}={\cal Z}^2_J R^{\pm}/{\cal Z}^{\pm}_{11}$,
are reported in Table~\ref{tab:lattices}. Further technical details 
will be provided in a forthcoming publication.

The values in Table~\ref{tab:lattices} show that $R^{\pm ,\; \mathrm{bare}}$
exhibit a pronounced mass dependence, which is more marked in
$R^{- ,\; \mathrm{bare}}$. We have explored several fit strategies, 
attempting to minimize the systematic uncertainties due to neglected higher orders in ChPT.
The structure of Eqs.~(\ref{eq:Reps}) and (\ref{eq:Rp}) indeed suggests
that it is possible to cancel large NLO ChPT corrections by constructing suitable combinations
of $R^{\pm ,\; \mathrm{bare}}$. We observe that the product $g_1^+\,g_1^-$ is very robust with respect to the details of the fit strategy. 
 The simplest 
way to extract this quantity is from a fit to the combination
$(R^+R^-)^{\mathrm{bare}}$, where NLO ChPT  corrections cancel in the limit $m\rightarrow 0$. We obtain
\be
\label{eq:prodLEC}
(g_1^+\, g_1^-)^\mathrm{bare} = 1.47(12)\, .
\ee
To extract $g_1^{+,\, \mathrm{bare}}$ and $g_1^{-,\, \mathrm{bare}}$ separately we then fit $R^{+ ,\; \mathrm{bare}}$ to NLO ChPT, 
taking the value of $F$ from a fit to the two-point functions
as in  Ref.~\cite{Giusti:2004yp} and the bare $\Sigma$ from Ref.~\cite{Giusti:2003rm}. Putting the result together with Eq.~(\ref{eq:prodLEC})
we get
\be
g_1^{+,\, \mathrm{bare}} = 0.63(4)(8)\, , \qquad g_1^{-,\,\mathrm{bare}} = 2.33(11)(30) \,,
\ee
where the first error is statistical and the second is an estimate of the 
systematic uncertainty from the spread of the central values obtained
from fits to different quantities and/or mass intervals.
\begin{figure}[t!]
\begin{center}
\includegraphics*[width=6.0cm]{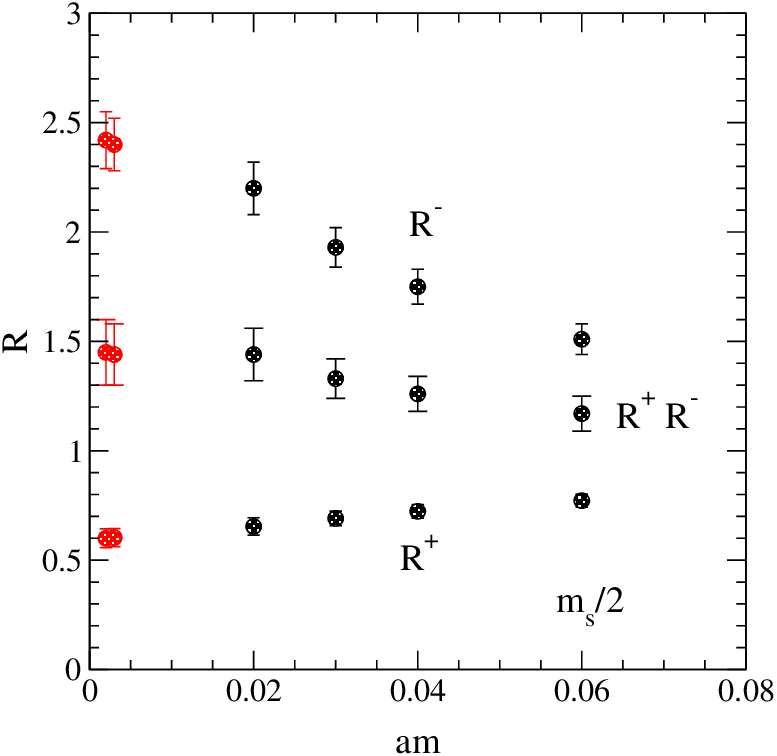}
\caption{Mass dependence of $R^{\pm, \mathrm{bare}}$ and $(R^{+}\cdot R^{-})^\mathrm{bare}$.
\label{fig:Rpm}}
\end{center}
\end{figure}
The physical LECs are given by
\be
g^\pm_1 =  k_1^\pm \left[\frac{R^{\pm, \mathrm{RGI}}}{R^{\pm, \mathrm{bare}}}\right]_\mathrm{ref} 
g_1^{\pm,\,\mathrm{bare}} \, ,
\ee
where $k_1^\pm$ are the renormalization group-invariant (RGI) Wilson 
coefficients~\cite{Gaillard:1974nj,Altarelli:1974ex,Altarelli:1980fi,Buras:1989xd,Giusti:2004an}. The RGI quantities
\be
 R^{\pm, \mathrm{RGI}}_\mathrm{ref} \equiv 
 R^{\pm, \mathrm{RGI}}\Big|_{r_0^2M_P^2=r_0^2M_K^2} 
\ee
at the pseudoscalar mass $r_0^2M_K^2=1.5736$ are taken from
Refs.~\cite{Guagnelli:2005zc,Palombi:2005zd,Dimopoulos:2006dm,Dimopoulos:SU4match}, and 
$r_0$ is a low-energy reference scale widely used in
quenched QCD computations~\cite{Necco:2001xg}. This
procedure, analogous to the one proposed for the scalar density in Ref.~\cite{Hernandez:2001yn},
provides values of the LECs that are non-perturbatively renormalized, 
as explained in detail in Ref.~\cite{Dimopoulos:SU4match}.
\vspace{-0.5cm}

\section{Physics discussion \label{sec:res}}
\vspace{-0.25cm}
By using the non-perturbative renormalization factors in 
Ref.~\cite{Dimopoulos:SU4match}
\be
\left[\frac{R^{+, \mathrm{RGI}}}{R^{+, \mathrm{bare}}}\right]_\mathrm{ref} = 1.15(12)\, , \qquad
\left[\frac{R^{-, \mathrm{RGI}}}{R^{-, \mathrm{bare}}}\right]_\mathrm{ref} = 0.56(6),
\ee
and the perturbative values $k_1^+=0.708$ and $k_1^-=1.978$ 
(see Ref.~\cite{Giusti:2004an}), we obtain our final results
\be
\label{eq:gpmfinal}
g^+_1 = 0.51(9)\,, \;\;\;  g^-_1 = 2.6(5)\, , \;\;\;  g^+_1 g^-_1 = 1.2(2)\; .
\ee
A solid estimate of discretization 
effects would require simulations at several lattice spacings, which is 
beyond the scope of this exploratory study. However, computations 
of $R^{\pm}$ at different lattice spacings and for masses close 
to $m_s/2$~\cite{Garron:2003cb,Giusti:2004an}
indicate that discretization effects may be smaller than 
the errors quoted above. It is also interesting 
to note that quenched computations of various physical quantities 
carried out with Neuberger fermions show small discretization 
effects at the lattice spacing of our 
simulations~\cite{Wennekers:2005wa,Babich:2005ay}.

The values of $g_1^{\pm}$ in Eq.~(\ref{eq:gpmfinal}) are the main 
results of this paper. They reveal a clear hierarchy 
between the low-energy constants, $g_1^{-}\gg g_1^{+}$, which implies
the presence of a $\Delta I=1/2$ enhancement in the GIM-limit
of (quenched) QCD. The strong mass dependence of $R^{\pm ,\; \mathrm{bare}}$ 
 in Fig.~\ref{fig:Rpm} 
indicates that an extrapolation of data around or above the physical 
kaon mass to the chiral limit is probably subject to large systematic uncertainties.

When the charm mass $m_c$ is sufficiently heavier than the three 
light-quark masses, the chiral effective theory has a 
three-flavour SU(3) symmetry and the LO $\Delta S =1$ effective 
Hamiltonian has two unknown LECs, $g_{27}$ and 
$g_8$. In our strategy these LECs are considered functions of the charm mass, and 
our normalizations are such that~\footnote{In the literature different normalizations 
of the LECs are used, e.g.  
${\rm g}_{\mbox{\tiny\underline{27}}}=(3/5)\, g_{27}$ and 
${\rm g}_{\mbox{\tiny\underline{8}}} = g_8/2$ in Ref.~\cite{Hambye:2003cy,Hernandez:2004ik}.}
\be\label{eq:gpmfinalsu3}
g_{27}(0) = g_1^{+}\, ,  \qquad g_{8}(0) = g_1^{-} + \frac{g_1^{+}}{5}\; .
\ee
The values of $g_{27}({\overline m}_c)$ and 
$g_{8}({\overline m}_c)$ can be estimated at the physical value of 
the charm mass ${\overline m}_c$ by matching the LO CHPT expressions 
with the experimental results for $|A_0|$ and $|A_2|$. The result is 
\be\label{eq:gpmexp}
|g_{27}^{\mathrm{exp}}({\overline m}_c)| \sim 0.50 \;, \qquad |g_{8}^{\mathrm{exp}}({\overline m}_c)|\sim 10.5\;. 
\ee
These estimates are, of course, affected by systematic errors due to higher-order 
ChPT contributions~\cite{Kambor:1991ah}. Keeping this in mind, the value of 
$g_{27}^{\mathrm{exp}}({\overline m}_c)$ is in good agreement with 
our result. Since $g_{27}$ is expected to have a mild dependence on the 
charm-quark mass (only via the fermion determinant in the effective gluonic action), and
barring accidental cancellations among quenching effects and higher-order 
ChPT corrections, this agreement points to the fact that higher-order ChPT 
corrections in $|A_2|$ may be relatively small. Our value for $g_8(0)$ differs by roughly 
a factor of 4 from $g_{8}^{\mathrm{exp}}({\overline m}_c)$ given in Eq.~(\ref{eq:gpmexp}).  
Apart from possible large quenching artefacts, 
our result suggests that the charm mass dependence and/or higher-order effects in ChPT are 
large for $|A_0|$. Indeed in this case penguin contractions, 
which are absent in the GIM limit, can be responsible for a large charm-mass 
dependence in $g_{8}$, a dependence that can be studied in the next step 
of our strategy~\cite{Giusti:2004an,Hernandez:2004ik}.

\vspace{-0.5cm}

\section*{Acknowledgements\label{sec:ack}}
\vspace{-0.25cm}

We are indebted to M. L\"uscher and P. Weisz, whose participation
in the early stages of this project was instrumental to our study. 
We would like to thank them also for many illuminating discussions. 
Our calculations were performed on PC clusters at CILEA, DESY Hamburg 
and the Universities of Rome ``La Sapienza'' and Valencia-IFIC, as well as on the IBM MareNostrum at the Barcelona Supercomputing Center, and the IBM 
Regatta at FZ J\"ulich.
We thankfully acknowledge the computer resources and technical
support provided by all these institutions and the University of 
Milano-Bicocca (in particular C. Destri and F. Rapuano) for its support.
P.H. acknowledges partial support from projects FPA2004-00996, 
FPA2005-01678 and GVA05/164.
\vspace{-0.75cm}

\bibliographystyle{apsrev}
\bibliography{DI12_SU4q}
\end{document}